\documentclass{sigchi}

\CopyrightYear{2020}
\setcopyright{acmcopyright}
\doi{https://doi.org/10.1145/3313831.3376866}
\isbn{978-1-4503-6708-0/20/04}
\conferenceinfo{CHI'20,}{April  25--30, 2020, Honolulu, HI, USA}
\acmPrice{\$15.00}

\usepackage{balance}       
\usepackage{graphics}      
\usepackage[T1]{fontenc}   
\usepackage{txfonts}
\usepackage{mathptmx}
\usepackage[pdflang={en-US},pdftex]{hyperref}
\usepackage{color}
\usepackage{booktabs}
\usepackage{textcomp}

\usepackage{microtype}        
\usepackage{ccicons}          

\usepackage{todonotes}
\usepackage{enumitem}
\usepackage{array}

\newcommand{\modified}[1]{\textcolor{black}{#1}}
\newcommand{\name}{DFSeer}

\def\plaintitle{{\name}: A Visual Analytics Approach to Facilitate Model Selection for Demand Forecasting}

\def\emptyauthor{}
\def\plainkeywords{
Interactive visualization; model selection; product demand forecasting; time series.}

\makeatletter
\def\url@leostyle{%
  \@ifundefined{selectfont}{
    \def\UrlFont{\sf}
  }{
    \def\UrlFont{\small\bf\ttfamily}
  }}
\makeatother
\def\pprw{8.5in}
\def\pprh{11in}

\definecolor{linkColor}{RGB}{6,125,233}
\hypersetup{%
  pdftitle={\plaintitle},
  pdfauthor={\emptyauthor},
  pdfkeywords={\plainkeywords},
  pdfdisplaydoctitle=true, 
  bookmarksnumbered,
  pdfstartview={FitH},
  colorlinks,
  citecolor=black,
  filecolor=black,
  linkcolor=black,
  urlcolor=linkColor,
  breaklinks=true,
  hypertexnames=false
}

\begin{document}

\title{\plaintitle}

\numberofauthors{1}
\author{%
  \alignauthor{Dong Sun\textsuperscript{1}, Zezheng Feng\textsuperscript{1}, Yuanzhe Chen\textsuperscript{2}, Yong Wang\textsuperscript{1}, Jia Zeng\textsuperscript{2}, Mingxuan Yuan\textsuperscript{2}, \\Ting-Chuen Pong\textsuperscript{1}, Huamin Qu\textsuperscript{1}\\
    \affaddr{\textsuperscript{1}The Hong Kong University of Science and Technology}\\
    \affaddr{\textsuperscript{2}Noah's Ark Lab, Huawei Technologies Co. Ltd.}\\
    \email{\textsuperscript{1}\{dsunae, zfengak, ywangct, tcpong, huamin\}@cse.ust.hk;  \textsuperscript{2}\{chenyuanzhe, zeng.jia, yuan.mingxuan\}@huawei.com}}\\
}

\maketitle

\begin{abstract}
Selecting an appropriate model to forecast product demand is critical to the manufacturing industry. 
However, due to the data complexity, market uncertainty and users' demanding requirements for the model, it is challenging for demand analysts to select a proper model. 
Although existing model selection methods can reduce the manual burden to some extent, they often fail to present model performance details on individual products and reveal the potential risk of the selected model. 
This paper presents {\name}, an interactive visualization system to conduct reliable model selection for demand forecasting based on the products with similar historical demand. 
It supports model comparison and selection with different levels of details. 
Besides, it shows the difference in model performance on similar products to reveal the risk of model selection and increase users' confidence in choosing a forecasting model. 
Two case studies and interviews with domain experts demonstrate the effectiveness and usability of {\name}. 
\end{abstract}




\section{Introduction}
In a manufacturing company, the forecast of product demand based on the historical demand data is critical to production planning and supply strategy developing \cite{gupta2003managing}.
Since there is often no model that is suitable for forecasting all the product demand \cite{ahmed2010empirical, fernandez2014we}, it has become increasingly important to select appropriate models for demand forecasting. 
An appropriate model for demand forecasting should be accurate, stable, and suitable for multiple products, which can optimize the allocation of production resources and reduce the risk of both order delay and the overstock of products \cite{gupta2003managing}. 

However, it is challenging to select a proper model for demand forecasting.
First, it is not easy to evaluate and compare different demand forecasting models due to the data complexity. The demand forecasting in a manufacturing company usually involves thousands of products and dozens of models. 
The domain experts often find it difficult to quickly evaluate and compare different demand forecasting models when the model performance is inconsistent on different products or in different time periods for the same product. 
Second, the market uncertainty, which is common in \modified{the} manufacturing industry, makes it even more difficult to find a reliable demand forecasting model \cite{lee2002aligning}. 
For example, a sudden decline in the price of substitute products may lead to a reduction in the demand, which can result in a large difference between the forecasted and exact demand.  
Third, it is challenging to find a model that fulfills the high requirements of product demand forecasting. The demand analyst is seeking a model that is accurate for most of the products, stable in the accuracy when forecasting the demand in different months, and applicable to as many products as possible. 

Some prior research has been conducted on the selection of forecasting models, which mainly focuses on 
parameter tuning \cite{bogl2013visual}, feature selection for modeling \cite{krause2014infuse, lu2014integrating}, and the comparison of different types of models \cite{feurer2015efficient, h2o2017automl, mljar2018automl, swearingen2017atm, wang2019atmseer, li2019manifold}. 
These studies provide helpful guidance for model selection and reduce lots of manual efforts in model selection. However, they often only provide the overall performance of models on the dataset and fail to reveal the performance details on individual products, which is important for decision making in demand forecasting model selection. 
For example, \textit{how accurate is the product demand forecast by different models?
How different is the demand forecasting performance of a certain model on different types of products (or different products of the same type)?
Are there some models that are suitable for the demand forecasting of most products?}
In addition, these studies do not reveal the risk of using a specific demand forecasting model. 
Instead of only getting a ranked list of models, users are also curious about the reliability of the models in forecasting the demand of the products with similar historical demand data, which is crucial to their choice of an appropriate demand forecasting model \cite{buono2007similarity, alvarez2010energy}.
Employing a high-risk model can lead to a reduction in revenue, loss of customers, and users' distrust of the model. 

In this paper, we propose {\name}, an interactive visual analytics system to support the comparison and selection of demand forecasting models in the manufacturing industry. 
Considering that a model with a good average performance on all the products may produce a poor forecast for some products, {\name} supports model selection according to the forecast of the products with similar historical demand data \cite{buono2007similarity, alvarez2010energy, faloutsos1994fast}. 
Our approach allows users to compare and select demand forecasting models through three levels of details, i.e., an overview, a product view, and a detail view. It can help users easily identify the models whose forecasting results are inaccurate or unstable for some products, as well as the models that are only suitable for a small number of products. 
Besides, the system combines the model performance with the historical demand data to reveal the risk of selecting a specific demand forecasting model.
It allows users to remove dissimilar products identified by integrating their domain knowledge with the properties of the time series extracted by automated algorithms. 
By showing the difference in model performance on the products with similar historical demand, {\name} can help users understand the possibility of an accurate forecast by the selected model, which can increase their confidence in their choice of the model. 
To the best of our knowledge, {\name} is the first visual analytics system to support the interactive selection of different types of time series forecasting models. 

We summarize the major contributions of the work as follows: 
\setlist{nolistsep}
    \begin{itemize}[noitemsep]
        \item A new workflow that supports demand forecasting model selection based on the products with similar historical demand and reveals the risk of selecting a model to increase users' confidence in adopting the model. 
        
        \item The implementation of a visual analytics system that enables users to compare and select demand forecasting models with different levels of details. 
        
        \item \modified{Detailed evaluations of the visualization system, including two case studies with real-world product demand data and interviews with four domain experts from a manufacturing company.} 
    \end{itemize}
\setlist{}
\section{Related Work}
Previous studies related to our work can be divided into three parts: techniques for time series forecasting, model selection for forecasting analysis, and visualization of discrete time series. 

\subsection{Techniques for Time Series Forecasting}
Recently a number of techniques have been proposed and applied to the forecast of time series, which can be categorized into three groups according to previous studies \cite{faloutsos2018forecasting, shi2018machine}: classical techniques for time series modeling, machine learning techniques, and deep learning techniques. 

Classical techniques for time series modeling mainly includes the autoregressive integrated moving average (ARIMA) model \cite{box2015time} and the exponential smoothing (ES) model \cite{hyndman2008forecasting}. 
The ARIMA model \cite{box2015time} computes the difference between two adjacent values in a time series to convert non-stationary data to stationary ones and utilizes the accumulated error terms to reduce the random fluctuation in time series forecasting. 
The ES model \cite{hyndman2008forecasting} emphasizes the effect of the recent time series on the forecast, without ignoring the effect of the time series from a long time ago. 
Machine learning techniques have also been applied to time series forecasting. 
For example, the k-nearest neighbor regression (KNN) model \cite{altman1992introduction} forecasts the future value of a time series based on its neighboring time series. 
Other machine learning techniques 
include the support vector regression (SVR) model \cite{gunn1998support}, gradient boosting \cite{friedman2001greedy}, 
Lasso \cite{tibshirani1996regression}, 
ridge regression \cite{hoerl1970ridge} 
and so on.
In addition, inspired by the excellent performance of deep learning techniques \cite{lecun2015deep}, 
researchers have also introduced deep learning techniques to time series forecasting. 
For instance, Model-Agnostic Meta-Learning (MAML) \cite{finn2017model} can be trained quickly even though there are only a small number of data points. 
When forecasting the future values of a time series, the recurrent neural network (RNN) \cite{mikolov2010recurrent} takes both the current value and past ones into consideration. 

Each of these techniques performs well on certain forecasting tasks, but their performance often declines on other forecasting tasks~\cite{ahmed2010empirical}. 
In this paper, we build an interactive visual analytics system to facilitate \modified{the} fast selection of demand forecasting models and help users easily identify 
the risk of a model. 

\subsection{Model Selection for Forecasting Analysis}
Various studies have been conducted to facilitate the selection of forecasting analysis models. These studies mainly consist of two types: visualization of forecasting models and automated approaches for model selection. 

Visual analytics techniques play an important role in helping users understand forecasting models. 
For example, Krause et al. \cite{krause2016interacting} and Hohman et al. \cite{hohman2019gamut} studied the influence of a feature on the forecast by varying the value of the feature. 
Visual analytics approaches have also been employed to explain the mechanism of deep learning models \cite{liu2016towards, ming2017understanding, wang2018ganviz}. 
Some other studies focus on using visualization techniques to facilitate the comparison and selection of forecasting models. 
For instance, TiMoVA~\cite{bogl2013visual} provides a detailed comparison of ARIMA models with different parameter settings to seek the best model. 
INFUSE~\cite{krause2014infuse} visualizes the significance of various features in forecasting analysis with a glyph design, where users can select different features to train a model and compare the generated models. 
ATMSeer~\cite{wang2019atmseer} exploits visualization techniques to support the interactive exploration of \modified{a} model selection process. 
\modified{Techniques in the statistics literature, such as the residual plot \cite{cook1982residuals} and the quantile-quantile (Q-Q) plot \cite{wilk1968probability}, have also been employed to evaluate and select models. The worm plot \cite{buuren2001worm} supports parameter tuning and the selection of LMS models \cite{cole1992smoothing}. Gabry et al. \cite{gabry2019visualization} introduced visualization techniques to the building of Bayesian models to help model developers evaluate and compare models.}
Apart from these visualization methods, some automated approaches for model selection have also been proposed.
For example, Auto-WEKA 2.0 \cite{kotthoff2017auto} and H2O AutoML \cite{h2o2017automl} are developed to help data scientists tune the parameters of different algorithms, compare the explored models, and find the model that performs the best on the dataset specified by the user. 

Although these techniques provide useful guidance for model selection and reduce the workload to some extent, they are not applicable to model selection for demand forecasting for two reasons. 
First, these approaches are aimed at the optimized performance of the whole dataset, which may result in an inaccurate forecast of the demand for some products. 
Second, they lack the performance details of models on individual products, which is important to decision making in demand forecasting model selection. 
Our work is proposed to solve the aforementioned problems, which supports demand forecasting model selection with different levels of details based on the forecast of the products with similar historical demand. Besides, it reveals the risk of the selected model by showing the performance difference on similar products. 

\subsection{Visualization of Discrete Time Series} 
Visual analytics approaches are widely used to explore time series~\cite{muller2003visualization, bach2014review}.
According to the survey by Aigner et al.~\cite{aigner2011visualization}, the studies on time series data visualization mainly includes three parts:
ordinal, discrete, and continuous time series visualization.
We will only discuss the prior work on visualizing discrete time series, which is more related to our approach.

One of the most popular visualization techniques for discrete time series is the line chart \cite{playfair1801commercial}, which describes how data change over time by connected line segments. An alternative of the line chart is the horizon graph \cite{saito2005two}, a space-saving technique that encodes the data by both the color and the height. 
Besides, the spiral \cite{weber2001visualizing} and calendar-based visualization \cite{van1999cluster} can reveal the periodic features of time series. To support the exploration of large scale time series, plenty of visual analytics tools have been developed. For example, PlanningVis \cite{sun2019planningvis} is a multi-level visualization system that juxtaposes heatmaps, line charts and bar charts to assist factory planners in optimizing a production plan. 

Inspired by the prior studies, we propose novel visualization techniques to 
show the discrete time series data and support the selection of demand forecasting models. 
\section{System Requirements and Design}
In this section, we interviewed four domain experts to understand the problem of model selection for product demand forecasting. Based on the comments and suggestions collected from the interviews, we summarized five requirements to guide the design of the visualization system. 

\subsection{Interview Study}
In order to understand the workflow of model selection for product demand forecasting, identify existing challenges in the approach currently used, and seek opportunities to improve the process of model selection, we conducted interviews with four domain experts from a manufacturing company. 
Two of the domain experts are algorithm developers who apply and train various models to forecast the demand of thousands of products, while the other two are demand analysts who are required to compare the performance of the models provided by algorithm developers and select the most suitable model. 

The interview for each domain expert lasted for one hour, which can be divided into three parts: the introduction to demand forecasting, the description of the workflow of model selection, and the discussion of challenges and user requirements. 
In the first part, the domain expert introduced the demand forecasting problem, the demand data used, and the algorithms to forecast product demand. 
In the second part, we asked the domain expert questions about the workflow of model selection for demand forecasting, including \textit{what features will be used to evaluate a model, what weights will be assigned to these features to rank the models, what information will be browsed by the domain experts}, and so on. 
In the third part, we discussed with the domain expert the problems in the current approach, the existing challenges, their requirements on a new approach for model selection, and the opportunities that visual analytics techniques can bring about. 

During the interview, the domain experts complained about the difficulty in model selection due to the large number of data they needed to explore. The current approach presents the forecast accuracy of each model for all the products in a tabular form, which is often overwhelming and requires a long scroll to explore the data. 
Also, the domain experts commented that they were not sure whether the selected model could produce an accurate forecast or not. 
In addition, they agreed with us on supporting model selection based on the products with similar historical demand. They commented that it follows the rule of product lifecycle \cite{stark2015product}, and hoped that the new approach can make full use of previous forecasting experience. 

\subsection{Design Requirements}
Based on the feedback collected from the interview study, we summarized the following five design requirements. 

\textbf{R1. Allow interactive model ranking.} 
The system should support the interactions that allow users to adjust the weights of different features for ranking demand forecasting models. 
In addition, it should also enable users to manually remove dissimilar products identified based on the features of the historical demand data extracted by automated algorithms. 

\textbf{R2. Enable model selection based on the forecast for similar products.} 
The system should allow users to interactively select the products with similar historical demand and help them find the models with the best performance on these products. This can take advantage of previous forecasting experience and alleviate the problem that a model selected according to the average performance on all the products may produce an inaccurate forecast for some products. 

\textbf{R3. Present the performance of models on different products.} 
Besides the overall performance of models on the selected products, the system should also show the performance of models on individual products, which is helpful for users to identify the drawback of a model and seek alternative models. 

\textbf{R4. Reveal the risk of an inaccurate forecast.} 
When applying a demand forecasting model, users have no idea whether the forecast is reliable or not. Revealing the risk of an inaccurate forecast can prevent the adoption of an unreliable model and increase the confidence in model selection. 

\textbf{R5. Support multi-level exploration.} 
Displaying the large quantity of data related to demand forecasting \modified{on} a limited screen is overwhelming and can result in visual clutter. The system should support the exploration of demand forecasting data with levels of details to reduce the manual effort required. 

\section{DFSEER} 

In this section, we introduce {\name}, an interactive visual analytics system that supports model selection for demand forecasting and reveals the risk of the selected model. 
This section first gives an overview of {\name} and then describes each component of the visualization system in detail. 

\subsection{System Overview}
\begin{figure}
\centering
  \includegraphics[width=\columnwidth]{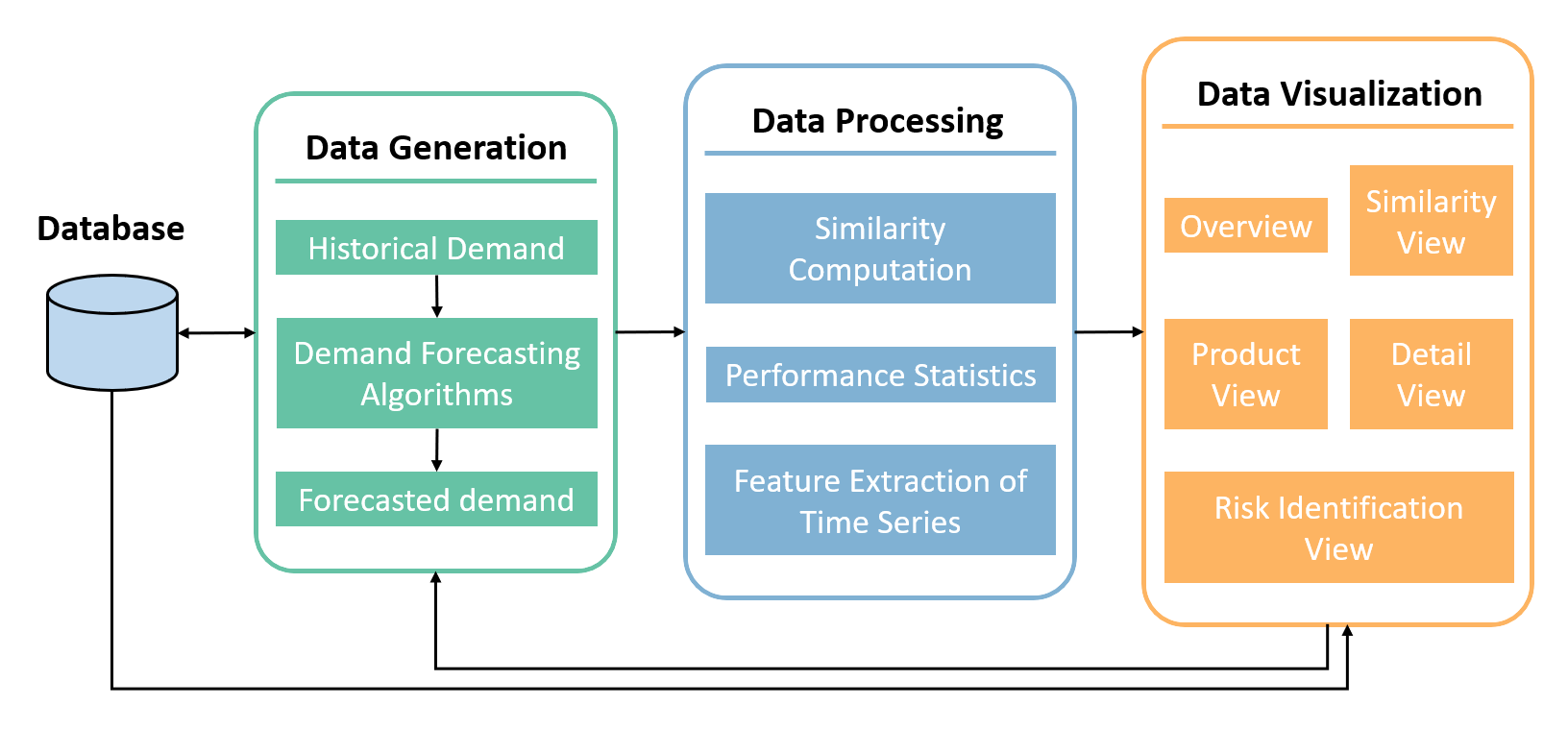}
  \caption{An overview of the DFSeer system, which mainly consists of three modules: data generation, data processing and data visualization.}
  \label{fig:system_overview}
\end{figure}

Figure~\ref{fig:system_overview} presents an overview of the {\name} system, which can be divided into three modules: data generation, data processing, and data visualization. 

The data generation module forecasts the demand for products in the next month with various kinds of forecasting algorithms. 
Three analytical tasks are then performed in the data processing module. 
First, the Euclidean Distance \cite{faloutsos1994fast} is utilized to measure the similarity between the z-normalized \cite{goldin1995similarity} historical demand data. 
Second, we compute the performance indicators to evaluate and rank different demand forecasting models. These performance indicators include the accuracy of the forecast in the next month, the variance of the forecast accuracy in the recent ten months, and the number of products to which a demand forecasting model is applicable. To be more specific, a model is applicable to a product when it is among the most accurate k models, where k is defined by users. 
Third, automated algorithms are introduced to extract the properties of the historical demand data, including the trend, the seasonality, the autocorrelation, and the stationarity. 
The demand data and the information generated by the data processing module are then sent to the visualization module. 

There are five components in the data visualization module: 
(a) \modified{an overview which allows rerunning the algorithm to forecast the demand in another month, 
supports interactively adjusting the weight for ranking demand forecasting models, 
and summarizes the performance of different models on the products selected by users}, 
(b) a similarity view which encodes the similarity between time series with the distance between the points in a scatter plot and allows users to select clusters for detailed analysis, 
(c) a product view which displays the detailed performance of models on individual products with a novel glyph design, 
(d) a detail view which presents the historical demand data of a user-specified product and the accuracy of the forecasts in past months by the top k models, 
and (e) a risk identification view which shows the features of similar historical demand data, and assists users in evaluating the risk of a model. 
The visualization module also supports several interactions such as linked-filtering and the tooltip.

With {\name}, the user can compare and select different types of demand forecasting models with levels of details. Besides, the methodology applied in model selection (i.e., model selection is based on the performance of models on the products with similar historical demand data) can increase the confidence of users in the selected model. 

\subsection{Ranking Models Interactively}

{\name} enables users to adjust the weight of the performance indicators for ranking demand forecasting models (\textbf{R1}). The models are re-ranked according to the adjustment, with the summarized performance indicators of each model presented in a ranked list. This presents the macroscopic information to support model selection for demand forecasting (\textbf{R2}, \textbf{R5}). 

\begin{figure*}
  \centering
  \includegraphics[width=1.85\columnwidth]{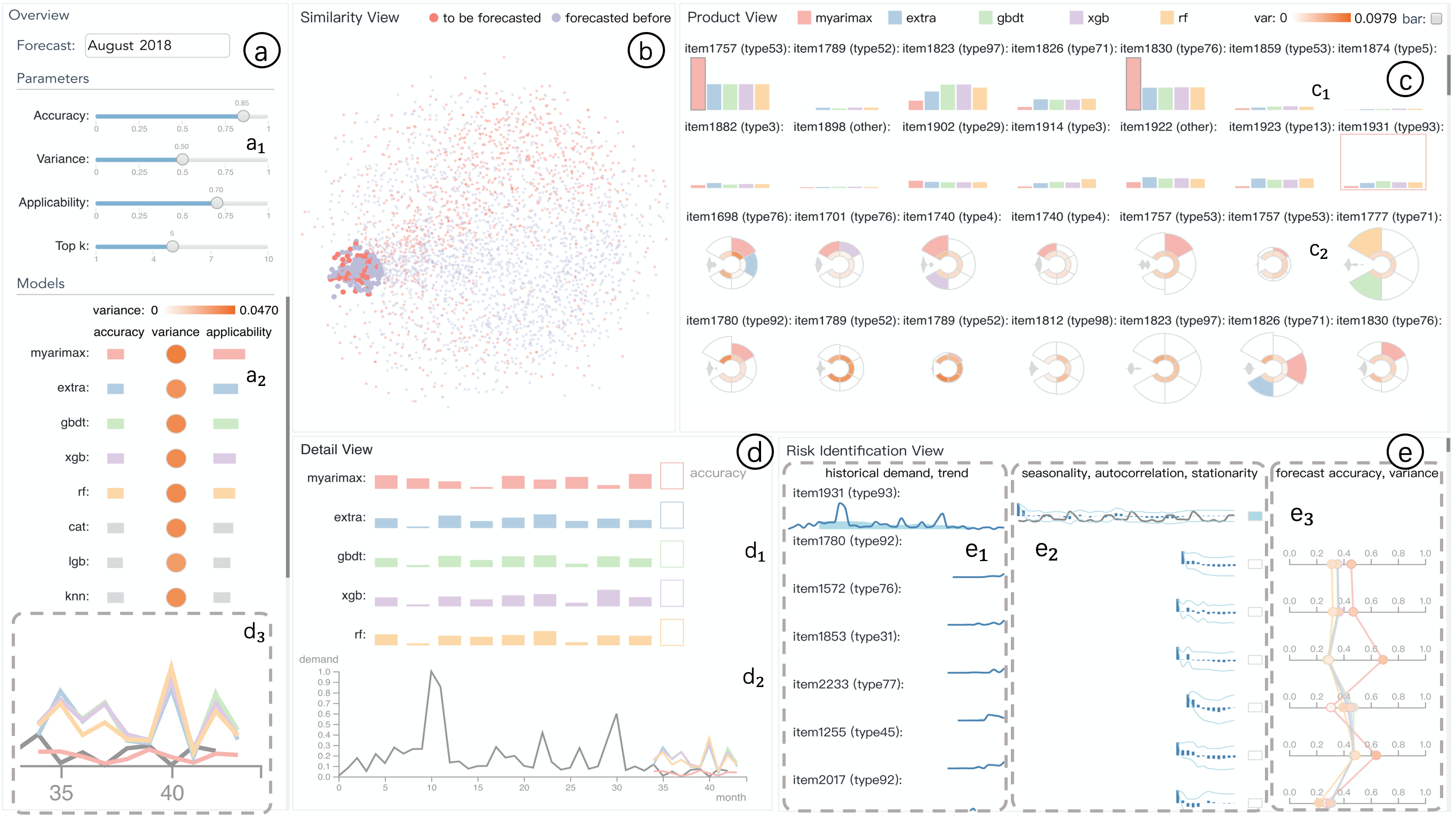}
  \caption{DFSeer supports interactive model selection for demand forecasting and reveals the risk of the selected model. 
  (a) The overview enables users to adjust the parameters for ranking demand forecasting models ($a_1$) and presents the ranked list of models ($a_2$). 
  (b) The similarity view allows users to select the products with similar historical demand data in a scatter plot. 
  (c) The product view encodes the model performance on individual products with bar charts ($c_1$) and carefully-designed glyphs ($c_2$). 
  (d) The detail view shows the monthly forecast accuracy of the top k models ($d_1$), and the real and forecasted demand of a user-selected product ($d_2$). The right part of $d_2$ is enlarged and shown in $d_3$. 
  (e) The risk identification view presents the extracted properties of the historical demand ($e_1$, $e_2$) and the model performance difference on similar products to reveal the risk of the selected model. 
  }~\label{fig:case_study_1}
  \vspace{-1em}
\end{figure*}

As shown in the upper part of the overview (Figure~\ref{fig:case_study_1}$a_1$), three sliders are provided for users to adjust the weight of the forecast accuracy, the variance of the forecast accuracy in recent months, and the number of products to which a demand forecasting model is applicable. 
There is another slider for parameter ``top k'', which indicates that among the models forecasting the demand of a product, only the most accurate k models \modified{are} considered applicable to that product. 
The ranked list of the demand forecasting models is shown below the sliders (Figure~\ref{fig:case_study_1}$a_2$), which summarizes the model performance on the products selected by users. For each model, the leftmost and rightmost rightward bars display the accuracy of the forecast and the number of the products to which a demand forecasting model is applicable respectively. \modified{The circle in the middle encodes the variance of the forecast accuracy in recent months with the color saturation, which is consistent with other views}. Only the top k models in the ranked list are depicted by different colors other than gray. 

\subsection{Selecting Similar Products}

The similarity view (Figure~\ref{fig:case_study_1}$b$) allows users to select the products with similar historical demand data in a scatter plot (\textbf{R2}). 
We employ the Euclidean Distance \cite{faloutsos1994fast} to measure the similarity between the z-normalized \cite{goldin1995similarity} time series. 
The Euclidean Distance is adopted because it is efficient \cite{ding2008querying} and can handle noisy data. Then, multidimensional scaling (MDS) \cite{borg2003modern} is applied to map the time series to a scatter plot, where the relative distance between two points indicates the similarity between two time series. In the scatter plot, the red points represent the time series to be forecasted, while the light purple points encode the time series forecasted before. Users can select a group of adjacent points by drawing a lasso to explore the performance of models on individual products. 

\subsection{Exploring Model Performance on Products}

After selecting a cluster of time series of interest, users can view the performance of models on individual products and further explore the forecast details of a user-specified product in recent months (\textbf{R3}). This enables users to select demand forecasting models by comparing their mesoscopic and microscopic information (\textbf{R2}, \textbf{R5}). 

\begin{figure}
  \centering
  \includegraphics[width=\columnwidth]{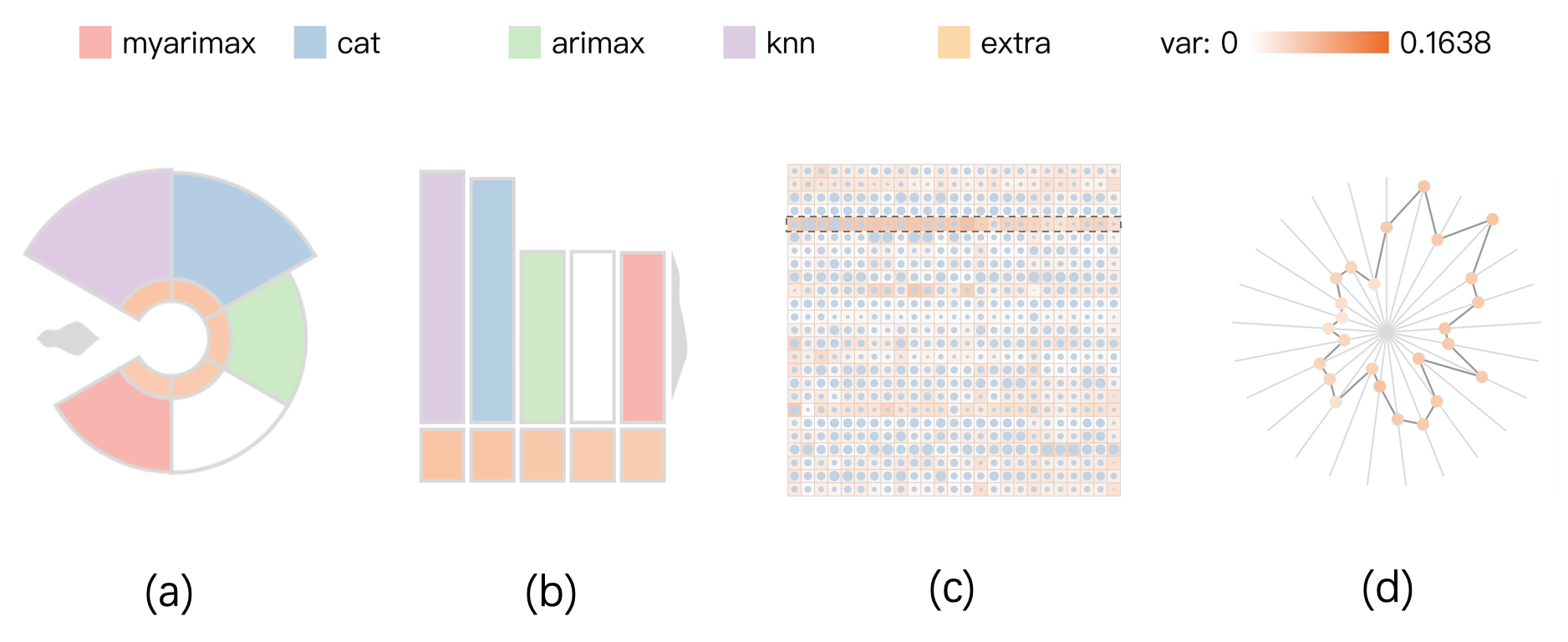}
  \caption{
  The glyph designs to describe the model performance on each product: 
  (a) the circular glyph design used in {\name}, 
  (b) a bar chart-based glyph design, 
  (c) a matrix view-based glyph design, 
  and (d) a star plot-based glyph design. 
  }~\label{fig:system_screenshot}
  \vspace{-1.2em}
\end{figure}

For the time series to be forecasted, 
the product view (Figure~\ref{fig:case_study_1}$c$) employs a bar chart (Figure~\ref{fig:case_study_1}$c_1$) to display the forecasted value of the top k models in the overview. The order of the bars is the same as that in the ranked list. 
For the time series forecasted before, we exploit a well-crafted glyph design (Figure~\ref{fig:case_study_1}$c_2$) to describe the performance of the forecasts by different models.
\modified{The current visual design is inspired by prior studies on multi-dimensional data visualization \cite{zhao2017skylens, zhao2011kronominer, chen2016dropoutseer, wang2018towards} and we worked closely with domain experts to collect their feedback on the visual design in an iterative manner.}
The glyph takes a circular design, which encodes the most accurate k models by different sectors. \modified{The sectors are arranged clockwise according to the forecast accuracy.} In each sector, the inner arc depicts the variance of the forecast accuracy in recent months with the color saturation, while the outer arc represents the accuracy of the forecast by the radius. The color of the outer arc encodes the model type. 
In the left part of the glyph, a violin plot shows the accuracy distribution of other models.
\modified{However, the circular glyph design is not very convenient for comparing model performance on different products. 
{\name} also supports a bar chart-based glyph design (Figure~\ref{fig:system_screenshot}$b$\modified), where the upper bars depict the forecast accuracy of the top k models and the lower rectangles encode the variance of the forecast accuracy with the color saturation. A marginal density plot on the right shows the accuracy distribution of other models.
Since the circular glyph design can help users distinguish the performance indicators on the products forecasted before (Figure~\ref{fig:case_study_1}$c_2$\modified) from the forecasting results of future product demand (the bar chart shown in Figure~\ref{fig:case_study_1}$c_1$\modified), and it is also visually attractive, {\name} uses the circular glyph design as the default choice. But users can interactively switch to the bar chart-based glyph design.
}

Two alternatives of the glyph are considered in our iterative design process, as shown in Figure~\ref{fig:system_screenshot}. 
Our earliest attempt is to use a matrix view (Figure~\ref{fig:system_screenshot}$c$) to describe the performance of models on individual products. Each row of the matrix represents a product, while each column represents a model. However, much manual effort is required to find the most suitable models for each product. 
Next, a star plot (Figure~\ref{fig:system_screenshot}$d$) is adopted to show the performance of different models, where each star denotes a product, and each spoke denotes the performance of a model. However, showing dozens of spokes is overwhelming and it is also not easy to identify the most accurate k models for each product in the star plot. 

After users click a bar chart (i.e., the time series to be forecasted) in the product view, the detailed performance of the top k models, the actual demand data and the latest forecast in the past months will be shown in the detail view (Figure~\ref{fig:case_study_1}$d$). 
\modified{Similar to the overview, a bar chart (Figure~\ref{fig:case_study_1}$d_1$) is used to express the accuracy of the forecasts in the past months.} The horizontal axis of the bar chart encodes the month, while the height encodes the accuracy. The bar charts are vertically aligned, whose order and color are consistent with those in the overview. 
In addition, users can also inspect the raw data (i.e., the actual and forecasted demand) in a multi-line chart below the bar charts (Figure~\ref{fig:case_study_1}$d_2$). 

\subsection{Identifying the Risk of the Selected Model}
When users select a model to forecast the demand, it is critical to reveal the risk of the selected model. \modified{The risk identification view (Figure~\ref{fig:case_study_1}$e$) is proposed to extract the properties of the time series (i.e., the historical demand of products) that are similar to the one that users specify in the product view and present the performance differences of the top k models on these time series.} It can help users identify the probability that the selected model will produce an accurate forecast (\textbf{R4}). 

To find the properties of a time series that are important for the forecast, we study how classical time series models make a forecast \cite{hyndman2007automatic, de2011forecasting} and discuss with domain experts how they judge whether two time series are similar.
We then identify four key properties of a time series, i.e., the trend, the seasonality, the autocorrelation, and the stationarity. 
To extract the properties of a time series, we first apply an additive model \cite{cleveland1976decomposition} to decompose a time series to get the trend and seasonality components. Later, the Autocorrelation Function (ACF) plot \cite{brockwell2016introduction} is utilized to analyze the correlation between the demand in different months. 
Additionally, the augmented Dickey-Fuller (ADF) test \cite{cheung1995lag} can reveal the stationarity of a time series.

In the risk identification view, the first row presents the properties of the time series to be forecasted, which is specified by users. \modified{The following rows show the properties of the time series forecasted before and the performance of the top k models, which are ranked by the similarity between the time series and the user-specified one.} 
In the left part of the risk identification view (Figure~\ref{fig:case_study_1}$e_1$), a line chart shows the historical demand, overlaid with an area chart to display the trend of the time series. 
\modified{Figure~\ref{fig:case_study_1}$e_2$ utilizes a black line to exhibit the seasonality of a time series}, which is overlaid with a bar chart to present the autocorrelation function. The two light blue lines indicate the 95\% confidence interval of the autocorrelation.
\modified{Besides, a rectangle is placed in the right part of Figure~\ref{fig:case_study_1}$e_2$ to encode the p-value output by the ADF test.} The rectangle is green if the historical demand data is stationary, or white if the data is not stationary. 
A vertically arranged parallel coordinates plot in the right part of the risk identification view (Figure~\ref{fig:case_study_1}$e_3$) shows the performance of the top k models in forecasting the demand of the similar products. 
In the parallel coordinates plot, each horizontal axis represents a time series forecasted before, \modified{while each of the top k models is denoted as a polyline. The intersection of the polyline and the axis is displayed as a circle, whose horizontal position encodes the forecast accuracy and color saturation encodes the variance of the forecast accuracy in recent months.} 
\modified{We encode the variance with the color saturation instead of the size of a circle or a bar to avoid possible occlusion.} 
When exploring the time series forecasted before and the performance of the top k models, users are allowed to remove a time series if they think it is dissimilar to the time series to be forecasted. 

\begin{figure}
\centering
  \includegraphics[width=\columnwidth]{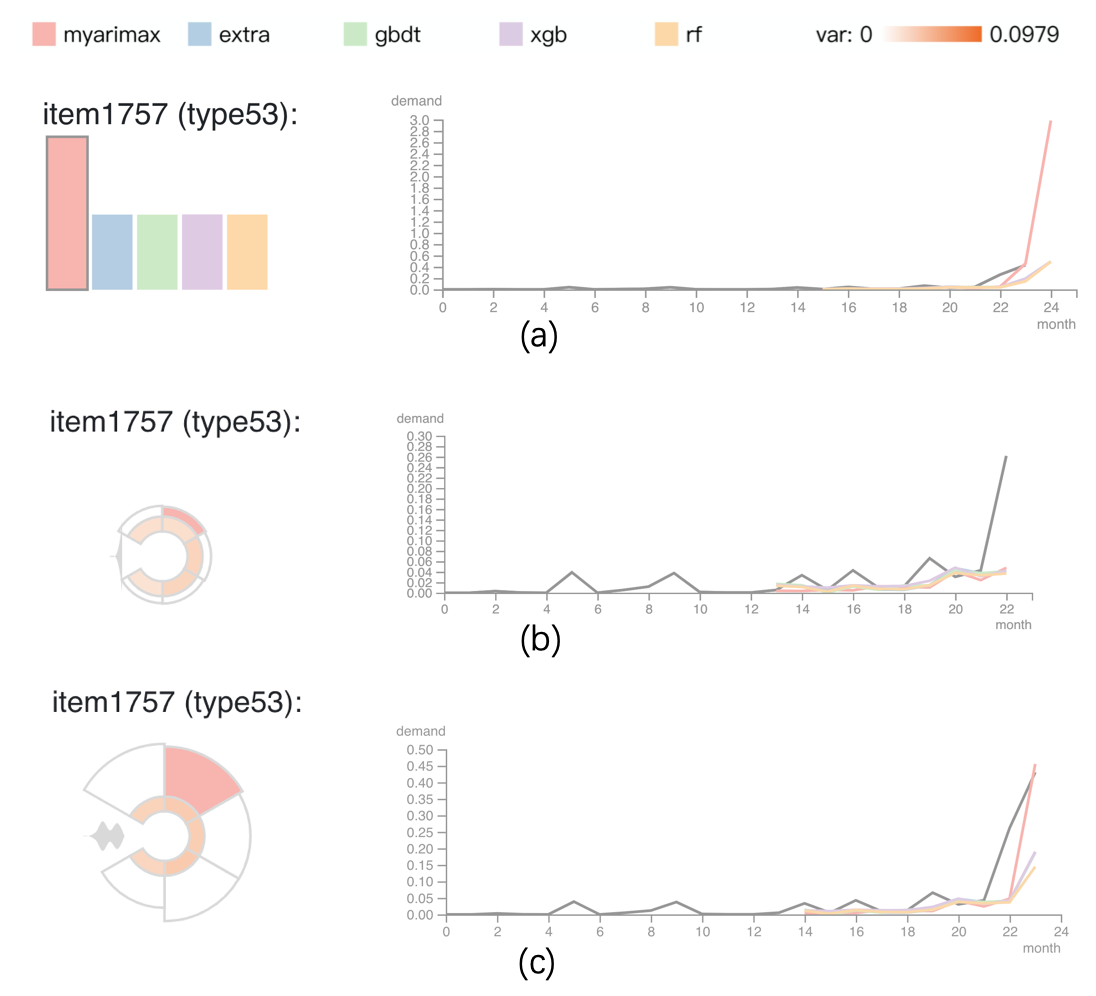}
  \caption{
  (a) \textit{item1757} gains a significantly high demand forecasting value by \textit{myarimax} in August 2018. 
  (b) An abrupt sharp increase in demand in June 2018 results in low forecast accuracy. 
  (c) \textit{myarimax} achieves a much better demand forecasting accuracy in July 2018 than other top models, indicating its excellent sensitivity in learning an abrupt demand increase. 
  }
  \label{fig:case_study_1_item_1757}
\end{figure}

\section{Case Study}

In this section, we conducted two case studies to demonstrate the effectiveness of {\name}.
The users involved in the case studies are the domain experts from the manufacturing company we collaborate with, who also attended the interview study in the next section. 
The dataset used 
consists of monthly demand of 1,096 products from January 2015 to November 2018. The actual names and types of the products are removed due to privacy issues.


\subsection{Inspecting Top Models: Best Model Does Not Work for All}

\begin{figure*}
  \centering
  \includegraphics[width=1.9\columnwidth]{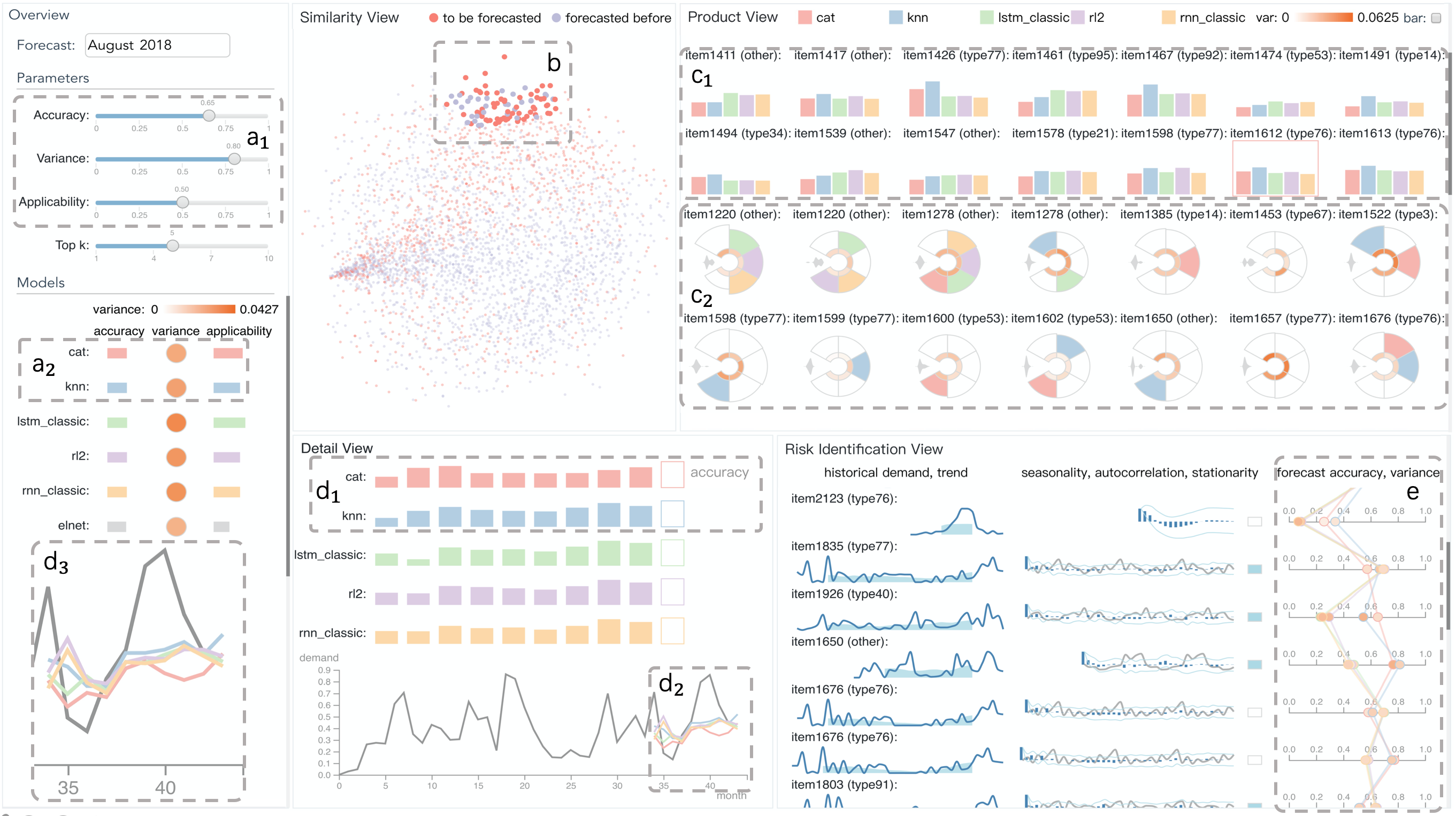}
  \caption{Identify the difference between ``similar'' models by using {\name}: 
  The user found that the top 2 models (i.e., \textit{cat} and \textit{knn}) have very similar average performance ($a_2$) and an overall similar distribution of ranking among the top 5 models for the selected products ($c_2$). He then clicked an example product of \textit{type76} (i.e., \textit{item1612}) and found the two models have similar variance of the monthly accuracy ($d_1$) and a close trend of the forecast accuracy for the similar products ($e$). But the forecasted demand by \textit{cat} is consistently lower than that by  \textit{knn} ($d_2$ or $d_3$). To mitigate the shortage of this type of key products, he selected \textit{knn} to forecast the demand for the products of \textit{type76}. 
  }~\label{fig:case_study_2}
  \vspace{-1em}
\end{figure*}

In the first case study, a user was asked to select a suitable model to forecast the demand in August 2018 for any cluster of products of his interest.
After loading the data to {\name} and setting the forecast period, he immediately found a cluster from the similarity view (Figure~\ref{fig:case_study_1}$b$).
The user first set the parameter \textit{top k} to 5
in the overview and filtered the data to gain a rough idea of the model performance (\textbf{R5}).
From the ranked list of models, the user quickly noticed that the forecast accuracy of different models is a bit low. Therefore, he further increased the weights of accuracy and applicability to check which model can lead to higher average accuracy (\textbf{R1}). The ranked list (Figure~\ref{fig:case_study_1}$a_2$) shows that the model \textit{myarimax} (e.g., a manually tuned ARIMAX model \cite{williams2001multivariate}) achieves the highest average accuracy. Meanwhile, the circular glyphs in the product view (Figure~\ref{fig:case_study_1}$c_2$) also indicate that \textit{myarimax} consistently ranks as one of the top 5 models when forecasting the demand of the majority of the similar products shown in the product view (\textbf{R3}). 
By quickly clicking the similar products to explore the performance details of the top 5 models, the user observed that \textit{myarimax} is more accurate than other top models when forecasting the demand in different months (Figure~\ref{fig:case_study_1}$d_1$) and the forecasted demand by the model is closer to the real one (Figure~\ref{fig:case_study_1}$d_2$ or \ref{fig:case_study_1}$d_3$). 
He also compared the performance of the top 5 models on similar products in the risk identification view and found that \textit{myarimax} is more accurate than other top models for most products (Figure~\ref{fig:case_study_1}$e_3$). 
Thus, the user became more confidence in \textit{myarimax} and decided to consider \textit{myarimax} as a candidate model for forecasting the demand of most of the selected products (\textbf{R2}). 



However, the user also quickly identified some abnormal forecasts from the product view. The red bars of two products (i.e., \textit{item1757} and \textit{item1830}) are surrounded by a black border, as shown in Figure~\ref{fig:case_study_1}$c_1$. 
It means that the product demand forecasted by \textit{myarimax} exceeds a threshold predefined based on domain knowledge. 
So the user chose one of the two products (i.e., \textit{item1757}) for further analysis. 
By looking at the detail view, the user found that \textit{item1757} and other products of the same type (i.e., \textit{type53}) usually have small demand values, but only \textit{item1757} gains a significantly high demand forecasting value by \textit{myarimax} (Figure~\ref{fig:case_study_1_item_1757}a). Such an abnormal observation gives the user the impression that the forecasting result of \textit{item1757} by \textit{myarimax} may be problematic. 
By further checking the historical demand of \textit{item1757}, the user easily observed that \textit{item1757} has an abrupt sharp increase in demand in June (Figure~\ref{fig:case_study_1_item_1757}(b)) and July (Figure~\ref{fig:case_study_1_item_1757}(c)). \textit{myarimax} achieves a much better demand forecasting accuracy in July than other top models, indicating its excellent sensitivity in learning an abrupt demand increase. However, it overrates this trend and forecasts an abnormally high product demand of \textit{item1757} in August.
For the other product (i.e., \textit{item1830}), the user also observed similar issues. 

These observations from {\name} informed the user of the potential risk of using \textit{myarimax} to forecast product demand (\textbf{R4}).   
Although it is the ``best'' model in terms of the overall performance, it can also overrate a sudden increase in product demand, resulting in demand forecasting failures.




\subsection{Breaking the Tie: ``Similar'' Models Can Be Different}
In this case study, another user was asked to finish exactly the same task as the first case study. But he selected a different group of similar products by brushing in the similarity view, as is shown in Figure~\ref{fig:case_study_2}b.

By browsing the product view (Figure~\ref{fig:case_study_2}$c_1$ and \ref{fig:case_study_2}$c_2$), the user easily noticed that \modified{many products are belonging to \textit{type76}}. Since the products of \textit{type76} are the sub-components for producing other products and play a critical role in the whole supply chain of the factory, a demand forecasting model that is both accurate and stable will be preferred. Therefore, the user significantly increased the weight of accuracy and variance (Figure~\ref{fig:case_study_2}$a_1$) to favor the demand forecasting models that have good average forecast accuracy and also perform consistently well along the time (\textbf{R1}).
The ranked list of models (Figure~\ref{fig:case_study_2}$a_2$) shows that the model \textit{cat} (i.e., CatBoost \cite{dorogush2018catboost}) and the model \textit{knn} (i.e., the KNN model \cite{altman1992introduction}) rank the first and second, respectively. But the two models have very \textbf{similar} performance in terms of the average accuracy and variance, making it difficult to distinguish which model is actually more appropriate for the selected products.
Therefore, he further checked the performance distribution of the top five models on the selected products whose demand was forecasted before (\textbf{R3}, \textbf{R5}). 
But the user still found that an overall \textbf{similar} distribution of \textit{cat} and \textit{knn} models that rank within the top five models for the selected products, as shown in Figure~\ref{fig:case_study_2}$c_2$.

However, when the user continued to closely explore the product view, he quickly observed that \textit{cat} (the red bars) tends to always forecast lower future demand for the selected products (especially the products of \textit{type76}) than \textit{knn} (the blue bars), as shown in Figure~\ref{fig:case_study_2}$c_1$.
This phenomenon is beyond the expectation of the user, since both models achieved \textbf{similar} accuracy and variance for different products.
Therefore, he selected a product of \textit{type76} (i.e., \textit{item1612}) from the product view to inspect more details.
The detailed view shows that \textit{cat} and \textit{knn} have a very similar demand forecasting accuracy along the time for \textit{item1612} (Figure~\ref{fig:case_study_2}$d_1$), and their demand forecasting results for other products with similar historical demand share a close trend (Figure~\ref{fig:case_study_2}e). 
However, when the user carefully checked the line charts of the forecasting results in the detail view (Figure~\ref{fig:case_study_2}$d_2$ or \ref{fig:case_study_2}$d_3$), he clearly observed that the demand forecasting result by \textit{cat} is \textbf{consistently lower} than that \modified{by} \textit{knn} along the time regardless of the drop and rise of the actual demand (\textbf{R4}). 
Such an observation was also found for other products of \textit{type76}.

As mentioned above, the products of \textit{type76} are critical in the whole supply chain of the factory. The shortage of these products can bring serious trouble to the manufacturing of the factory. Therefore, despite the similarity between \textit{cat} and \textit{knn}, the user finally decided to choose \textit{knn} as the demand forecasting model for the products of \textit{type76}, since it often has higher demand forecasts than \textit{cat} (\textbf{R2}). Though such a choice can increase the risk of overstock, it can better mitigate the shortage of these key products.

\section{Expert Interview}

We conducted semi-structured interviews with four domain experts working on model selection for demand forecasting to demonstrate the effectiveness and usability of {\name}. This section first introduces the interview procedures. We then summarize the feedback collected from the experts. 

\subsection{Participants and apparatus}
The participants include four domain experts from the manufacturing company we work with. \modified{The domain experts are different from the experts attending the requirement analysis interview and haven't participated in the design of the {\name} system.} 
Two of the domain experts ($E_1$ and $E_2$) are algorithm developers who are working on developing demand forecasting algorithms, while the other two ($E_3$ and $E_4$) are demand analysts who are responsible \modified{for} selecting the most suitable demand forecasting models for different products from dozens of algorithms provided by algorithm developers. 
The interviews were conducted with a 23.8-inch 1920 x 1080 monitor showing the visual analytics system. 

\subsection{Dataset and Tasks}
The dataset used in the expert interviews is the same as that in the case studies, which contains the monthly demand data of 1,096 products from January 2015 to November 2018. 
During the interviews, we asked the domain experts to fulfill two carefully-designed tasks to assess the usability of the visualization system and understand how the domain expert selects a demand forecasting model. 
\setlist{nolistsep}
    \begin{itemize}[noitemsep]
        \item \textbf{T1.} Analyze the result of model selection for a group of similar products. 
        \item \textbf{T2.} Select suitable models to forecast the demand of pre-selected products. 
    \end{itemize}
\setlist{}
In \textbf{T1}, we plan to select a model to forecast the demand of products in August 2018 and pre-select a group of similar products. The corresponding model performance on these products is displayed in the visualization system. The domain experts are then asked to answer nine questions related to the model performance in different levels of details, as shown in Table~\ref{tab:task_1_questions}. In \textbf{T2}, the domain experts are required to find the best model for two types of products, i.e., the general products which emphasize more on the accuracy of a demand forecasting model (e.g., switches and indicator lights), and the products that are important to the supply chain of the factory and need a more stable model (e.g., high-speed optical modules and digital signal controllers). 

\begin{table}
  \centering
  \begin{tabular}{ m{1em} | m{21em} }
    \hline
    Q1 & Find the most accurate demand forecasting model. \\
    \cline{2-2}
    Q2 & Find the most stable demand forecasting model. \\
    \cline{2-2}
    Q3 & Find the demand forecasting model which has the best applicability. \\
    \hline
    Q4 & Find the model whose forecast for the target month is different from other top k models. \\
    \cline{2-2}
    Q5 & Find the product for whom the forecast by the top k models is the most accurate. \\
    \hline
    Q6 & For a given product, find the most unstable model among the top k models. \\
    \cline{2-2}
    Q7 & For a given product, find the model whose forecast is the closest to the real demand. \\
    \hline
    Q8 & Find the product whose historical demand is the most dissimilar to the selected one. \\
    \cline{2-2}
    Q9 & Given the products which are similar to a selected one, find the product whose forecasted demand is the most accurate. \\
    \hline
  \end{tabular}
  \caption{The questions in T1 are related to model performance in different levels of details: the overview (Q1 - Q3), the product view (Q4 - Q5), the detail view (Q6 - Q7), and the risk identification view (Q8 - Q9).}~\label{tab:task_1_questions}
  \vspace{-1em}
\end{table}

\subsection{Procedure}
During the interview, we first introduced the methodology of demand forecasting model selection, the visual design and interactions, and the functionality of each part of the {\name} system to the domain experts. 
After that, an example usage scenario was given to help them understand how to use the visualization system to select demand forecasting models. 
Then, the participants were allowed to explore the system on their own. 
The aforementioned tutorial lasted for about 20 minutes. 
After the tutorial, the domain experts were invited to discuss how they select a demand forecasting model and what information is used to support their decision in their daily work. 
Then, the participants were guided to carry out two tasks. 
Both tasks employed a think-aloud protocol, and the comments and suggestions of the domain experts were all recorded. 
After accomplishing the two tasks, the participants were asked to describe the workflow of using the {\name} system to select demand forecasting models. 
They were also encouraged to compare {\name} with the approach currently used in their daily work. 
Then, we invited the participants to rate the visual analytics system on a 7-point Likert scale (1 - strongly disagree, 7 - strongly agree) from three aspects. The detailed questions are shown in Table~\ref{tab:questionnaire}. 
Finally, the domain experts attended a post-study interview where they gave comments and suggestions on the workflow and trust, the visual design and interactions, and the usability of the {\name} system. 
The study for each domain expert lasted for approximately one hour and 20 minutes. 

\begin{table}
  \centering
  \begin{tabular}{ m{1.5em} | m{20.5em} }
    \hline
    Q1 & The model selection based on the products with similar historical demand is reasonable. \\
    \cline{2-2}
    Q2 & The workflow provides useful guidance for me to select suitable demand forecasting models. \\
    \cline{2-2}
    Q3 & The workflow can reduce my workload in model selection for product demand forecasting. \\
    \cline{2-2}
    Q4 & I have confidence in the model selected by using the visualization system. \\
    \hline
    Q5 & The visual design is intuitive. \\
    \cline{2-2}
    Q6 & The visual design is easy to understand. \\
    \cline{2-2}
    Q7 & The visual design provides enough information for me to select a demand forecasting model. \\
    \cline{2-2}
    Q8 & The visual design and interactions help me compare and select demand forecasting models. \\
    \hline
    Q9 & The visual analytics system is easy to use. \\
    \cline{2-2}
    Q10 & The visual analytics system is helpful for demand forecasting model selection. \\
    \cline{2-2}
    Q11 & I would like to use the visual analytics system to select demand forecasting models in the future. \\
    \cline{2-2}
    Q12 & I will recommend the visual analytics system to my colleagues working on forecasting model selection. \\
    \hline
  \end{tabular}
  \caption{The questionnaire consists of three parts: the workflow and trust (Q1 - Q4), the visual design and interactions (Q5 - Q8), and the usability (Q9 - Q12).}~\label{tab:questionnaire}
  \vspace{-1em}
\end{table}

\subsection{Results} 
We summarize the feedback of the interviews and the result of the questionnaire into three parts: 

\textbf{Workflow and trust.} 
All the domain experts agreed that the workflow for demand forecasting model selection in our work is helpful and increases their trust in the selected model. 
$E_3$ said ``In our daily work, 
model selection is based on the average accuracy of a model. It is difficult for me to trust that the model would give an accurate forecast for a certain product.'' 
He commented the proposed workflow to support the selection of models based on their performance on similar products is reasonable and trustful, since the algorithm will often produce similar results for similar time series data \cite{buono2007similarity, alvarez2010energy, faloutsos1994fast} and the change of demand over months follows the rule of product lifecycle \cite{stark2015product}. 
$E_2$ mentioned ``Having seen the detailed model performance on similar products, I have a better understanding of the models and trust the model I selected.''
$E_4$ reported it was difficult to manually identify the products with dissimilar historical demand. With the properties extracted and presented in the risk identification view, he was able to find dissimilar time series without much effort. 
\modified{When performing \textbf{T2}, $E_1$ found the model he selected is different from that by others. After checking the risk identification view, he realized many dissimilar products were included. He then shrunk the scope of selecting similar products and got a suitable model. 
$E_1$ commented ``The risk identification view helps me check whether the scope of selecting similar products is suitable.''}

The domain experts also gave suggestions on the selection of products with similar historical demand data. $E_2$ suggested involving the relationship between different types of products in the computation of similarity. $E_4$ mentioned there are too many points in the similarity view and recommended that the system can further filter products by the type of products and the relationship between them. 

\textbf{Visual design and interactions.} 
The domain experts reported that the visual design is informative and the interactions are user-friendly. 
$E_3$ 
compared {\name} with the approach currently employed in their work and 
commented ``{\name} provides an intuitive and easy way for me to explore the performance of models. Without it, I need to scroll a large tabular form to search for the information I need.''
$E_4$ liked the glyph design in the product view and mentioned ``The bar chart can reveal which model produces a different forecast, while the circular glyph shows which model is more accurate or stable.'' 
He also stated that the glyph design is in line with their daily work, where the domain experts only focused on several top-ranked models. 
\modified{The domain experts were surprised by the capability of the risk identification view. $E_1$ found there were several products with few historical demand records and removed them from the risk identification view. He commented ``the tool combines the power of automated algorithms and our knowledge to judge whether a product is similar or not''.} 
$E_2$ mentioned the interactions are easy and powerful. 
He said ``Through the interactions with the visualization system, including adjusting the weights for ranking models, selecting similar products, and removing dissimilar historical demand data, I get highly involved in how the models are ranked and gain a better understanding of the models.'' 

Despite the positive feedback, the domain experts commented that the product view and the risk identification view had a steep learning curve, especially for the users who have little knowledge of visualization. 

\textbf{Usability.} 
The domain experts confirmed that the {\name} system is easy to use and can facilitate their work in model selection for demand forecasting. 
\modified{It took them about 25 minutes to finish the tasks (including the task introduction), where no task was skipped. Only one participant made an error when answering the objective questions in \textbf{T1} (i.e., Q5), which is because there are two products with very similar model performance. This shows that {\name} helps analyze the model performance on similar products.} 
$E_4$ said ``Compared with our daily work where the similarity between products is manually determined based on the type of products, {\name} provides a more reasonable and flexible manner for finding similar products.''
The domain experts agreed that {\name} provides helpful guidance for them to find an appropriate demand forecasting model and lightens their workload a lot. 
$E_3$ mentioned the visual analytics system was easy to learn and use. He appreciated the nice experience of using the system, which was better than using the tedious tabular form. \modified{The domain experts planned to deploy the {\name} system in the manufacturing factory to facilitate their daily work of model selection. They agreed it is worth spending around 20 minutes learning to use the system.} 

Besides the positive comments on the usability of the {\name} system, the domain experts also put forward several constructive suggestions. $E_4$ expressed that the system could involve other advanced similarity calculation methods and allow users to choose which method to use. $E_1$ suggested there was a need \modified{for} post-processing which can provide a feasible demand value when the selected model offers an abnormal forecast. 

\section{Discussion}
Although {\name} has shown its ability in facilitating model selection for demand forecasting, limitations still exist. \modified{In this section, we discuss the limitations from the perspectives of visual design scalability, lessons learned, and generalizability and extensions.} 


\subsection{Visual Design Scalability} 
The similarity view of the {\name} system displays 3,288 time series for users to select similar products. To present a larger number of time series, measures shall be taken to reduce the overlap of data points. For example, we can adopt a hierarchical strategy which first asks users to assign the type of products and specify the range of distance for searching similar products, and then shows the products inside this range. 
The visual analytics system only presents the detailed performance of less than ten models. 
This is in accordance with the workflow of selecting demand forecasting models in the manufacturing factory, where domain experts only focus on several models with the best performance. Showing the detailed performance of all the models could increase the visual burden of users. 

\subsection{\modified{Lessons Learned}}
\modified{
By working with domain experts, we realize that showing the internal structure of models to demand analysts may not be that useful. In our study, the demand analysts often do not have a strong background in machine learning models. Hence, instead of showing the internal structure of models, {\name} visualizes the patterns of the input (i.e., time series), the output (i.e., forecasted demand and performance indicators), and the correlations between them. 
Besides, we find the model ranked first by the algorithm may not always be the best. For example, in our second case study, the domain experts selected the model ranked second because compared with the model ranked first, it can relieve the shortage of the products that are critical to the supply chain. 
In addition, our work shows that model transparency is highly appreciated and needed in model selection for demand forecasting. Instead of only being told to use a model, users also want to know why some time series are considered similar and how the model performance is on these similar products. 
} 

\subsection{Generalizability and Extensions}
\modified{Although in our study, {\name} is used to support model selection for demand forecasting, it can be adapted to other application scenarios, such as stock market prediction, power usage prediction, traffic flow forecast, and medical diagnosis. For example, in the scenario of power usage prediction, {\name} can group residents based on their residence locations and incomes, and help power managers find suitable prediction models for different groups of residents and identify potential power shortages. In the scenario of medical diagnosis, {\name} can be adapted to diagnosing the illness according to the cases with similar symptoms, which can help doctors make a diagnosis by summarizing the predictions of multiple models.} 
{\name} can also be extended to increase its power in model selection for demand forecasting. First, other competitive time series forecasting models can be involved in our system. Second, the relationship between different types of products can be considered when computing the similarity between time series. Third, the system can also provide more distance calculation strategies for users to choose from. 

\section{Conclusion and Future Work}
We have presented {\name}, a visual analytics system to help data analysts select demand forecasting models based on the products with similar historical demand. 
The visualization system provides different levels of details for users to explore and compare the performance of demand forecasting models. 
Besides, the {\name} system can reveal the risk of a model by showing the performance difference of the model on similar products. 
We evaluated the visual analytics system through two case studies with product demand data and semi-structured interviews with four domain experts working on model selection for demand forecasting in a manufacturing factory, which shows that {\name} is helpful and effective in facilitating model selection for demand forecasting. 

In future work, we plan to include more time series forecasting algorithms in the visualization system. Besides, we would like to consider the relationship between different types of products when computing the distance between the historical demand data. We also want to enable users to filter products by type in the similarity view. Finally, we intend to provide other advanced approaches to compute the distance between time series and allow users to select the approach on their own. 
\section{Acknowledgments}

We thank all the domain experts participating in the study and the reviewers for their constructive suggestions. This work is partially funded by RGC GRF 16241916. 

\balance{}

\bibliographystyle{SIGCHI-Reference-Format}
\bibliography{sample}

\end{document}